\documentclass[useAMS,usenatbib]{mn2e}
\usepackage{mathtext,amssymb}
\usepackage{epsfig}
\usepackage{graphics}
\usepackage{graphicx}
\usepackage{url}

\newcommand{\kms}{\ensuremath{\mbox{km~s}^{-1}}}
\renewcommand{\arcsec}[1]{^{\prime\prime}\!\!\!#1\,}   
\renewcommand{\arcmin}[1]{^{\prime}\!\!\!#1\,}
\newcommand{\pc}{\ensuremath{\rm pc}}
\newcommand{\kpc}{\ensuremath{\rm kpc}}
\newcommand{\magdot}[1]{^{\rm m}\!\!\!#1\,}
\newcommand{\Msun}{\ensuremath{\mathrm{M_\odot}}}
\newcommand{\acknowledgements}{\subsection*{Acknowledgements}}

\renewcommand{\mag}[1]{^{\rm m}\!\!\!#1\,}

\newcommand{\href}[1]{\url{#1}}
\title{Spectral Study of the WNL Star FSZ35 in the M33 Galaxy}

\author[Maryeva and Abolmasov]{O.\ Maryeva $^{1}$\thanks{E-mail:
olga.maryeva@gmail.com} and P.\ Abolmasov$^{2}$\\
$^{1}$Stavropol State University, Pushkina str., 1, Stavropol, 355009, Russia \\
$^{2}$Sternberg Astronomical Institute, Moscow State University, Universitetsky pr., 13, Moscow, 119992, Russia}

\begin{document}

\date{Accepted -- . Received -- ; in original form --}

\pagerange{\pageref{firstpage}--\pageref{lastpage}} \pubyear{2011}
\maketitle
\begin{abstract}
        We study and analyse low-resolution spectra of the unordinary late
        WN star FSZ35 in M33. We classify the object as a hydrogen-rich WN8
        star. Using the radiative transfer code CMFGEN, we determine the
        physical parameters of this object and compare them to the
        parameters of other WN8 stars including the LBV star Romano's star
        during the minimum of brightness. Unlike Romano's star, the object
        is fairly stable both spectrally and photometrically, that may be
        attributed to its more advanced evolutionary stage or lower luminosity.
        FSZ35 is shown to possess a
        compact nebula producing faint but detectable [O\,{\sevensize III}]
        emission. 
        Location of this object at a large distance ($\sim 100 \pc$) from the
        nearest association suggests the object may be one more example of a
        massive runaway star. 
\end{abstract}

\begin{keywords}
   galaxies: individual: M33 -- stars: Wolf-Rayet -- stars: supergiants -- stars: individual: FSZ35 (M33)
\end{keywords}
\section{Introduction}
        Since middle 1990s, radiation transfer codes designed for modeling
        extended expanding atmospheres became a reliable instrument for
        studying hot stars with high mass-loss rates. 
        These codes make it possible to calculate 
        the atmospheres of outstanding
        objects like O-supergiants, Wolf-Rayet (WR) stars and even Luminous Blue
        Variables (LBV). In particular, the CMFGEN \citep{Hillier5} 
        code was used to model the spectra of $\eta$~Car \citep{grohetacar}, 
        AG~Car \citep{groh} and AFGL2298 \citep{Clark09}. 
        CMFGEN and PoWR \citep{Hamann} were used to determine the properties  
        of the majority of the Galactic nitrogen-rich Wolf-Rayet (WN)
        stars \citep{Crowther1995,HamannPoWR} 
        including WN stars near the Galactic center \citep{Martins,HamannGC}.

        High-luminosity stars are rare objects, and their studies in our
        Galaxy are complicated by dust absorption, crowded stellar fields and
        unknown distances. Therefore, it is important, and sometimes more
        convenient, to study massive stars in nearby galaxies. 
        Studying WR stars that are probable progenitors of Type Ibc Supernovae 
        and gamma-ray bursts  is of especial importance (see review by \citet{Crowtherreview} 
        and references therein).
        It is even more difficult to determine the distances toward Galactic WN8 stars
        because they can not be associated with clusters or
        associations \citep{CrowtherWR,marchenko98}. 
        One possible scenario proposed to explain the apparently enigmatic characteristics 
        of WN8 stars is that they are runaways, either ejected via 
        dynamical interaction from the cores of forming dense star clusters (e.g. \citet{Evans}), 
        or accelerated by a slingshot-type mechanism during the supernova explosion of the initial primary 
        component in a binary system \citep{deDonder}. 

        The comprehensive catalogue of \citet{massey98} lists in total
        117 nitrogen-rich Wolf-Rayet stars in M33 (including WN candidates). 
        But models of  atmospheres are constructed for only few objects. 
        Parameters of the atmospheres of MCA1-B and B517 were
        determined with the iterative technique of \citet{Hillier90} 
        (precursor of CMFGEN) in \citet{smith} and \citet{b517}, respectively. 
        Using the WM-BASIC code \citep{WMBASIC}, \citet{MasseyM33} used the
        ultraviolet spectra of six known late WN (WNL) stars in M33 to 
        estimate their fundamental parameters: wind terminal velocities,  
        luminosities, effective temperatures, radii, mass-loss rates 
        and C/N abundance ratios. In the current work, we use the CMFGEN 
        code \citep{Hillier5} to study the spectrum of the little-studied 
        WN star FSZ35 in M33.

\begin{table*}\centering
\caption{Observational log for the SCORPIO data. 
S/N is signal-to-noise ratio, PA is position angle.}
\label{tab:obstabscor}
\bigskip
\begin{tabular}{lcccccccc}
\hline
             &  Exposure        &          & Spectral    &                  &      &                   & Spectral    &            \\
Date         &    time          & Grism    &   range     & $\delta \lambda$ & S/N  & Seeing            & standard    &  PA         \\
             &      [s]         &          &   [\AA]     &      [\AA]       &      & [$\arcsec$~]       & star        &  [$\degr$] \\ 
\hline
\multicolumn{9}{c}{\bf FSZ35}    \\
4. 10. 2007  &  $900+1200$      & VPHG1200G&  4000-5700  & 5.5               & 19   & 1.2               &  G191B2B   & -182      \\
             &                  &          &             &                   &      &                   &            &          \\
                           \multicolumn{9}{c}{\bf V532}  \\ 
5. 10. 2007  &  $3\times 900$   & VPHG1200G&  4000-5700  & 5.5               &  34  & 1.1               & BD25d4655  & -141     \\
\multicolumn{9}{c}{ }  \\
\hline
\hline
\end{tabular}
\end{table*}
       
            The object  FSZ35 ($\alpha=01^{h} 33^{m} 00^{s}\!.20$,     
            $\delta=+30^{o} 30\arcmin{}~ 15\arcsec{\ .}2$, (J2000 epoch)) is situated near 
            the association 128~OB\footnote{From the catalogue of OB
              associations by  \citet{HumphreysSandage}}
            in the M33 galaxy 
            14$\arcmin{\,}$ away from galactic centre. 
            Projected distance toward the association is 30-40$\arcsec{\,}$ or 100-120~pc. 
            Ivanov et al. were the first to obtain photometrical
            data for this object (IFM-B 174, in their notation) 
            in the {\it U}, {\it B}, and {\it V} bands and publish an identification chart for it \citep{Ivanov}. 
            The star is listed in the H$\alpha$ emission-line object
            catalogue \citep{fsz} as object number 35.
            In 1998, Massey \& Johnson 
            published a list of 22 Wolf-Rayet stars selected 
            by photometry in three non-standard filters: WN ($\lambda4686$), WC 
            ($\lambda4650$) and CT ($\lambda4752$, continuum).
            Identification was then confirmed spectroscopically \citep{massey98}.
            In this work, the star is listed as object E1 and is classified 
            as a Wolf-Rayet star of nitrogen sequence, WN8 subtype.
            \citet{Sholukhova} note similarity between 
            the spectra of  FSZ35 and V532 (Romano's star). The latter
            is a better studied object located in the outer spiral arm of M33 
             at a distance of about $17\arcmin{\,}$ \ from the centre.
             V532 is universally recognized as an LBV star. 
             It demonstrates pronounced photometrical and 
             spectral variability \citep{kurtev,polcaro,me}.
             Hence we pay especial attention to comparison between these two
             objects. Similar modeling for V532 will be published in a separate
             paper \citep{v532model}.  

             This paper is organized as follows. The observational data
             and data reduction process are described in the next section. 
             We characterise and classify the observed spectrum in 
             Section~\ref{sec:spclass}.
             In Section~\ref{sec:model} we describe the basic properties of the
             CMFGEN code and its main assumptions, while in Section~\ref{sec:res}
             we present and analyse the modeling results. 
             In Section~\ref{sec:disc} we discuss the results.
             Finally, in Section~\ref{sec:con} we summarize the main points of our work. 

\section{Observations and Data Reduction}\label{sec:obs}

              In this work, we use a spectrum of FSZ35 obtained at the 
              6m Special Astrophysical Observatory (SAO) 
              telescope\footnote{Spectral data were taken from the archive of 
              Special Astrophysical Observatory (SAO) of Russian Academy of 
              Sciences, 
              \href{http://www.sao.ru/oasis}}. \ 
              Two exposures, 1200s and 900s in length, were obtained with the SCORPIO multi-mode  
              focal reducer \citep{scorpio} in the long-slit mode on October 4, 2007. 
              VPHG1200~G grism was used providing the spectral range of 3950-5500~\AA.
              The data were reduced using the {\tt ScoRe} package 
              for long-slit data reduction, written in IDL
              especially for SCORPIO long-slit data reduction. 
              This package consists of procedures created by V.Afanasiev, 
              A.Moiseev, P.Abolmasov and O.Maryeva.
              Package includes all the standard stages of long-slit
              data reduction process. 
              The final spectrum has spectral resolution  $\sim
              5.5$~\AA\ (weakly dependent on wavelength) and  signal-to-noise
              ratio per resolution element in continuum $\sim 20$.

              Besides the spectrum of FSZ35, in this work we use a
              spectrum of Romano's star obtained at the 6m SAO telescope 
              practically at the same time with the same grating under 
              similar conditions. Details about the spectral data on V532 and 
              data reduction may be found in our recent paper \citet{me}. 
              Basic information about the observational data on both objects is 
              summarized in Table~\ref{tab:obstabscor}.

\section{Spectral classification}\label{sec:spclass}

\begin{figure*}
\centering
\epsfig{file=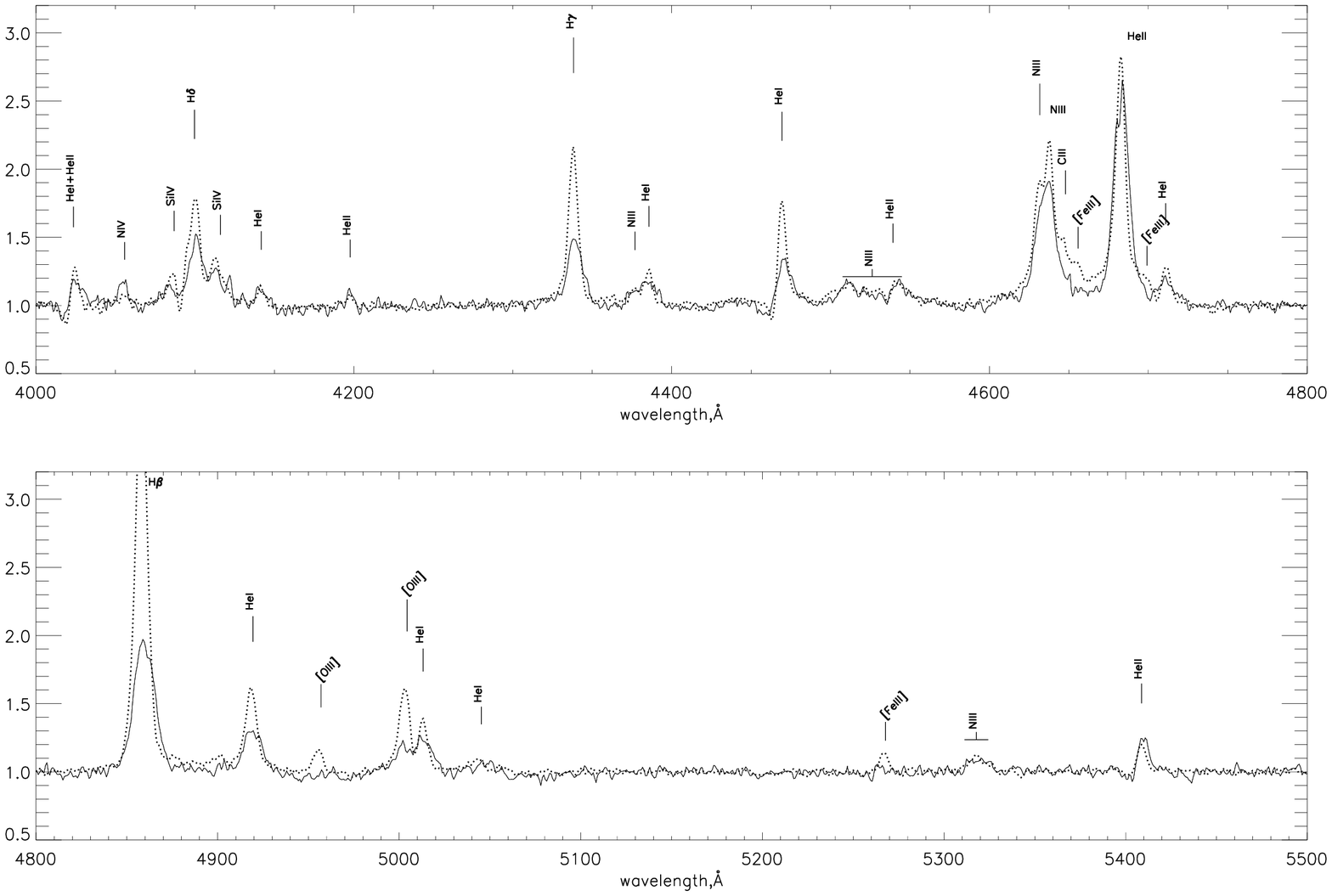,width =1.0\textwidth}
\caption{Spectra of  FSZ35 (solid line) and V532 (dotted line) 
         obtained  in October 2007. 
         The spectra are normalized by the local continuum level.}
\label{fig:fsz35v532}
\end{figure*}
          Figure~\ref{fig:fsz35v532} shows the spectra of V532 and FSZ35 obtained 
           under similar conditions. The spectral appearance of FSZ35 shows strong 
           similarities with V532. The   spectrum of FSZ35 is as rich as the spectrum 
           of V532 in emission lines, but the presence of nebular
           lines is questionable (see below Section~\ref{sec:disc})
           and hydrogen lines are fainter.
           Table~\ref{tab:linetab} presents the lines detected in the 
           spectrum of FSZ35. 
           Lines of helium and hydrogen in the spectrum of FSZ35  are much 
           broader in comparison with the these lines in the spectrum of V532. 
           For example, FWHM (Full Width at Half Maximum) of H$\beta$ is $12.0\pm0.2$ and
           $5.5\pm0.1$\AA\ for FSZ35 and V532, respectively. 
           For He\,{\sevensize I}$\lambda 4921$, FWHM are $9.5\pm0.2$ and $4.4\pm0.1$\AA.

\begin{table*}\centering
\caption{List of emission lines detected in the spectrum of FSZ35. For
  higher signal-to-noise ratios, equivalent widths (EW) are given. }
\label{tab:linetab}
\bigskip
\begin{tabular}{lcc||crr}
$\lambda$, \AA \  &    Ion             &    EW, \AA\            &
  $\lambda$, \AA\ &    Ion & EW, \AA\      \\
\hline
 3964.73          &  He\,{\sevensize I}               &                  &   4523.60         &  N\,{\sevensize III}  &    \\
 3970.08          &  H$\epsilon$+He\,{\sevensize II}  &                  &   4530.80         &  N\,{\sevensize III}  &    \\
 3994.99          &  N\,{\sevensize II}               &                  &   4534.60         &   N\,{\sevensize III} &    \\
 4009.00          &  He\,{\sevensize I}               &                  &   4541.60         &  He\,{\sevensize II}  &    \\
 4025.60          &  He\,{\sevensize I}+He\,{\sevensize II}          &                  &   4547.30         &   N\,{\sevensize III} &    \\
 4057.80          &  N\,{\sevensize IV}               & $1.10 \pm 0.15$  &   4601.50         &  N\,{\sevensize II}   &    \\
 4088.90          &  Si\,{\sevensize IV}              & $1.2 \pm 0.3$    &   4607.20         &  N\,{\sevensize II}   &    \\
\multicolumn{3}{l}{
$\left. \hspace{-0.2cm}
\begin{tabular}{lr}
 4097.31          &  ~~~~~N\,{\sevensize III}     \\  
 4101.74          &  ~~~~H$\delta$+He\,{\sevensize II}     \\
 4103.40          &  ~~~~~N\,{\sevensize III}     \\
\end{tabular}\  \hspace{0.8cm} \right\}4.4\pm 0.4$ }  &
\multicolumn{3}{|c}{ \hspace{-1.0cm}
\begin{tabular}{crr}
4613.90  &      ~~~~~~~~~~ N\,{\sevensize II}  &  ~~~~~~~   \\
4621.40  &     ~~~~~~~~~~  N\,{\sevensize II}  & ~~~~~~~   \\
4630.54 &    ~~~~~~~~~~   N\,{\sevensize II}  & ~~~~~~~     \\
\end{tabular}
 } \\
 4116.10          &  Si\,{\sevensize IV}              & $2.4 \pm 0.4$    &   4634.00                 &  N\,{\sevensize III}  & $  7.2\pm 0.1 $   \\
 4120.99          &  He\,{\sevensize I}               &                  &   4640.64                 &  N\,{\sevensize III}  & $  4.9\pm 0.1 $   \\
 4143.76          &  He\,{\sevensize I}               &$1.0 \pm 0.3$     &   4643.09                 &  N\,{\sevensize II}   &                  \\
 4199.80          &  He\,{\sevensize II}              &                  &   4650.16                 &  C III & $  3.3\pm 0.1 $   \\
 4236.93          &  N\,{\sevensize II}               &                  &   4685.81                 &  He\,{\sevensize II}  & $20.0 \pm 0.5$   \\
 4241.79          &  N\,{\sevensize II}               &                  &   4713.26                 &  He\,{\sevensize I}   &  $2.8 \pm 0.2$   \\
 4241.79          &  N\,{\sevensize II}               &                  &   4861.33                 &H$\beta$&  $13.4\pm 0.3$  \\
 4340.47          &   H$\gamma$+He\,{\sevensize II}   & $5.5 \pm 0.4$    &   4921.93                 &  He\,{\sevensize I}   & $4.1 \pm 0.2$   \\
 4387.93          &  He\,{\sevensize I}               &                  &{\footnotesize5001$\div$5007}&  N\,{\sevensize II}+[O\,{\sevensize III}]   &                 \\
 4471.69          &  He\,{\sevensize I}               &$3.0 \pm 0.3$     &   5015.67                 &  He\,{\sevensize I}   & $2.9\pm 0.2$   \\
$\left. \hspace{-0.0cm}
\begin{array}{l}
 4510.90      \\  
 4514.90       \\
 4518.20        \\
\end{array} \hspace{0.0cm}\
 \right\} $ & N\,{\sevensize III} &  $\sim 1$&
$ \left.
\begin{array}{l}
5314\\ 5320\\ 5327\\
\end{array} \right\}$ &  N\,{\sevensize III}  &    \\
& ~~~~~~~~~ & ~~~~~~~~~~~~~~ &5411.50         &  He\,{\sevensize II}  &  $1.8\pm 0.2$  \\
\hline
\end{tabular}
\end{table*}

            Using nitrogen and He\,{\sevensize II} lines, we classified V532 in October 2007 as a WN8 star
            using the classification of \citet{smithprinja} for WN6-11
            stars. Locations of V532 on the diagrams ``equivalent width of He\,{\sevensize I}~$\lambda 5876$
            versus He\,{\sevensize II}~$\lambda 4686$'' and  ``equivalent width of He\,{\sevensize II}~$\lambda 4686$ 
            versus FWHM of this line'' are consistent with the WN8 subtype \citep{me,polcaro10}. 

            Similar consideration allows to classify FSZ35 as a WN8 as
            well. Besides this, the N\,{\sevensize IV}$\lambda 4057$ line clearly seen in our
            spectrum is never present in WN9 spectra, thus excluding
            FSZ35 identification as a later-subclass object. 
            Unfortunately, our data cover only a limited spectral range
            from 3900 to 5500\AA. Therefore we can not use 
            the He\,{\sevensize I} $\lambda 5876$ line for spectral
            classification. 

\begin{figure}
\centering
\epsfig{file=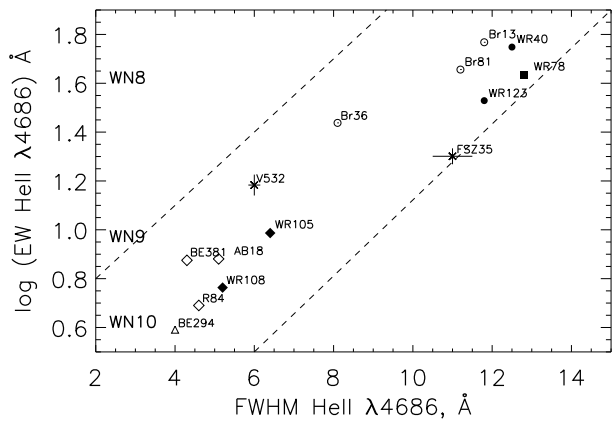,width =1.1\linewidth}
\caption{The location of FSZ35 and V532 (October 2007) on equivalent width versus FWHM 
         diagram for the ${\rm He\,{\sevensize II}}\lambda 4686$ line. Known Galactic (filled symbols)
         and LMC (open) WN stars are shown for comparison: WN7 by squares, 
         WN8 by circles, WN9 by diamonds, and WN10 by triangles. Data on these objects 
         were taken from  \citet{CrowtherLMC,CrowtherBohannan}. 
        }
\label{fig:spclass}
\end{figure}

            We used a quantitative chemistry-independent criterion based on the
            FWHM of the He\,{\sevensize II}~$\lambda 4686$ line for alternative spectral classification 
            (see, for example, \citet{smith}). 
            Equivalent width (EW) of the He\,{\sevensize II}~$\lambda 4686$ line in the spectrum of FSZ35 is 20\AA. 
            FWHM of this line is $10.5\pm0.5$\AA.
            In Fig.~\ref{fig:spclass}, we show the location of V532 and FSZ35 on the diagram
            of the EW of He\,{\sevensize II}~$\lambda 4686$ versus the FWHM of this line. 
            Position of the object in this diagram is consistent with
            our identification of FSZ35 as a WN8 star.

            We confirm the spectral classification of the object as a WN8
            star. This our result also suggests the object is much less variable
            than V532 that spans a broad range of effective temperatures and
            luminosities and becomes a late WN star only in deep minima. FSZ35 is
            thus closer to ``quiet'' WN stars.
 
 \section{Modeling}\label{sec:model}  
            
              For our analysis we used the non-LTE radiative transfer code
              CMFGEN \citep{Hillier5}.
              CMFGEN solves radiative transfer equation for objects with 
              spherically-symmetric extended outflows using either the Sobolev 
              approximation or the full comoving-frame solution of the
              radiative transfer equation. To facilitate simultaneous 
              solution of the transfer equation  and statistical 
              equilibrium equations, partial linearization method is used. 
              To facilitate the inclusion of metal line blanketing in CMFGEN, 
              superlevel approach \citep{Anderson89,Anderson91}  is used. 
              In this formalism, levels with similar 
              properties are treated as one and have identical
              departure coefficients. This allows to
              save considerable amount of computer memory and time. 
              Recent versions of the code incorporate also the effect of level dissolution, 
              influence of resonances on the photoionization cross section, 
              and the effect of Auger ionization.
       
              Clumping is incorporated into CMFGEN using volume
              filling factor approach \citep{Hillier99}. Filling factor is allowed to depend
              on radius. By default, the wind is considered homogeneous at the 
              hydrostatic radius and becomes more and more clumped with the wind velocity. 
              Clumping tends to reduce the derived mass-loss rates by a factor of 
              $\sim 3-5$ \citep{marcolino}.
              Unclumped mass loss rate (i. e. calculated not taking
              clumping into account) is related to the volume filling factor $f$ by 
              relation  $\dot{M}_{\rm uncl}=\dot{M}_{\rm cl}/\sqrt{f}$ (see
              for example \citet{HHH}).    

              Every model is defined by a hydrostatic stellar radius $R_*$, 
              luminosity $L_*$, mass-loss rate $\dot{M}_{\rm cl}$, filling factor $f$,  wind 
              terminal velocity $v_\infty$,  stellar mass $M$, and
              by the abundances $Z_i$ of included elementary species. 
              We assumed a constant turbulent velocity of 20 km/s. Its
              variations do not affect the resulting spectrum much
              save for the equivalent width of HeI$\lambda$4686 that
              becomes slightly brighter in more turbulent
              atmospheres. 
\section{Modeling Results}\label{sec:res}
        
           Using the model of Romano's star calculated by
           \citet{v532model} as the seed model we adjusted its parameters to reproduce the
           observed spectrum of FSZ35. As for V532, H, He, C, N, Si, Fe, O, Ne, 
           Mg, S, Ar, Ca, and Na were included in calculations.
           We considered only sub-solar iron abundances (between 0.2 and 0.5 solar).
           Hydrogen lines in the spectrum of FSZ35 are weaker than in the spectrum of V532, 
           suggesting lower H/He values of $\sim 0.6\div 0.8$ (by number), 
           unlike V532 where hydrogen abundance is estimated as
           H/He$\simeq$1.3. 

           We increased the effective temperature of the model in 
           order to reproduce the N\,{\sevensize IV}$\lambda 4057$ emission present 
           in the spectrum, and then varied the luminosity at a constant 
           effective temperature level to fit the observed V-band luminosity. 
           ${\it V}=18\magdot{\ .}7$ is reported by \citep{massey98}, in consistence 
           with the observed spectrum slightly hotter than that of V532 
           in deep minimum. In our calculations we supposed that distance toward M33 is $D=847\pm60 \ \kpc$
           and the distance modulus is $(m-M)=24.64\pm 0.15\magdot{\,}$ \citep{distance}. 
           We neglected intrinsic extinction of M33, since the object
           is distant from center of the galaxy.  Extinction in the
           Galaxy is estimated as ${\it E(B - V)\/} =
           0\magdot{\ .}029$ (corresponding to $A_v=0\magdot{.}1$ for
           $R_V=3.1$) by the NED
           extinction calculator \citep{schlegel98}. 
           
           At the next step, we change the mass loss rate and 
           terminal velocity $v_\infty$.
           Lower resolution and spectral range do not allow to make reliable
           estimates of the terminal wind velocity, as we did for V532
           in \citet{me}. 
           Terminal velocity given in Table~\ref{tab:parmodel} was found by
           fitting the observed line profiles. The filling factor was
           set to $f=0.1$ for all the models.
           At the last step of the modeling process, we adjust the
           nitrogen abundance. 

\begin{figure*}
\centering
\epsfig{file=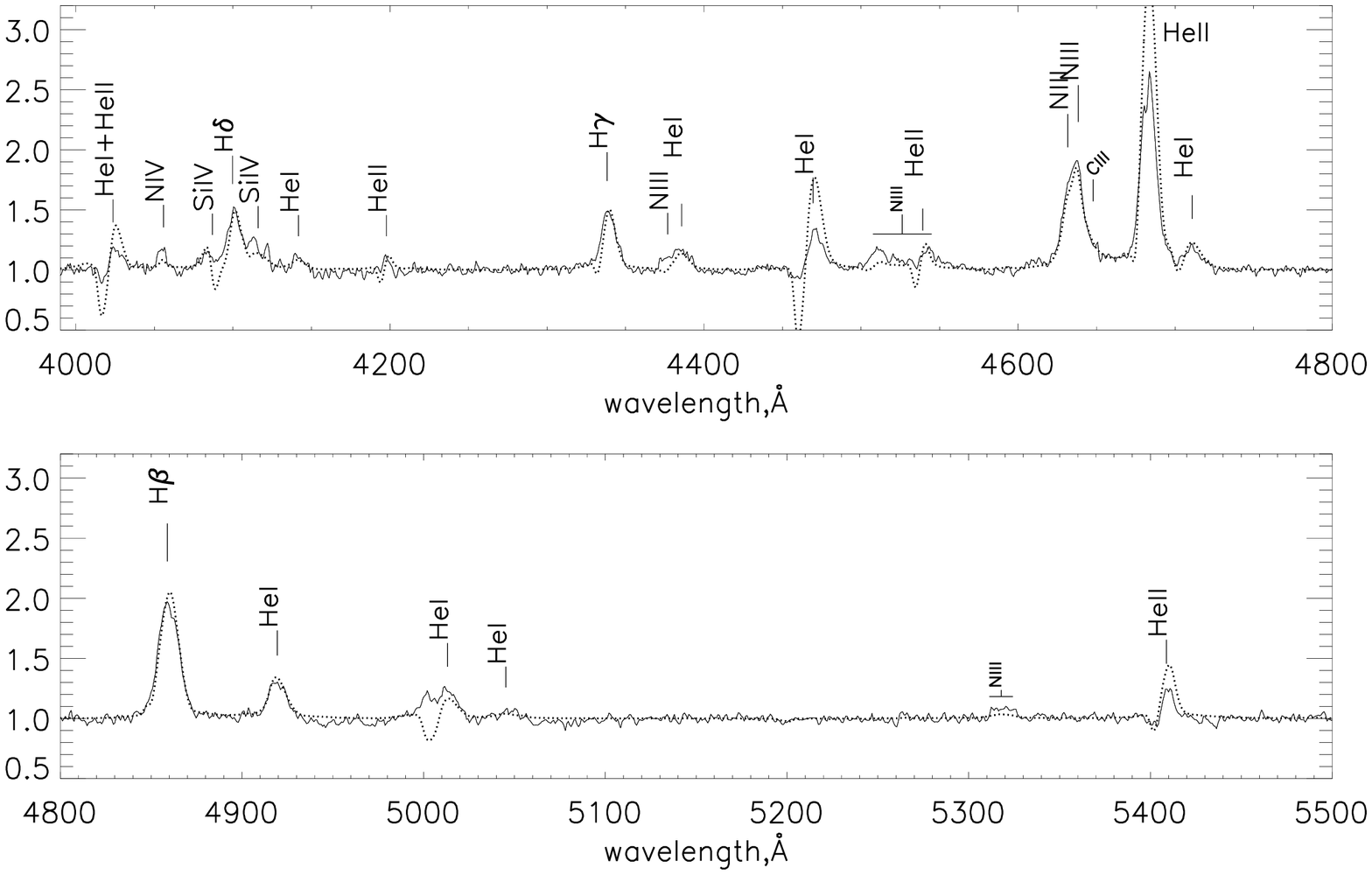,width =1\textwidth}
\caption{The spectrum of FSZ35 (solid line) compared with the best-fit CMFGEN model (dotted line).}
\label{fig:modelfsz35}
\end{figure*}

           In Figure~\ref{fig:modelfsz35} we present the spectrum of FSZ35 and our model. 
           We succeed in achieving good agreement with
           the observational equivalent widths and profiles of hydrogen lines and
           singlet lines of neutral helium as well as the line He\,{\sevensize I}$\lambda 4713$. 

           In the best-fit model, H/He ratio is 0.8 and, consequently, 
           mass fraction of hydrogen is $\simeq$0.17.
           FSZ35 conforms to the definition of H-rich WN stars \citep{Nugis}, 
           that must have hydrogen mass fraction $>0.05$. 
           The mass loss rate is $\dot{M}_{\rm cl}=2.6\cdot 10^{-5}\mbox{M$_{\odot}$/year}$ and 
           $\dot{M}_{\rm uncl}=8.22\cdot 10^{-5}\mbox{M$_{\odot}$/year}$.  
           The mass loss rate is given in Table~\ref{tab:parmodel} 
           together with  other wind and stellar parameters. 
         For comparison, the values of these parameters for some other WN stars
         taken from the literature are given in the table. 
         Note that the parameters of WR116 and WR89  were calculated
         using PoWR code \citep{Hamann} while the parameters of the other stars using CMFGEN. 
         Table~\ref{tab:parmodel} shows that the parameters of FSZ35 are similar 
         to these for some other WN8 stars given in the table, irrespective of a 
         code which was used the estimate the parameters. 
         In our calculations we neglected intrinsic extinction in M33. 
         If extinction near FSZ35 is $A_v=0\magdot{\ .}1$, 
         luminosity increases by about 10\% and becomes $L_*=6.3\cdot10^{5}L_{\odot}$.

             FSZ35 differs from  V532 in the values of mass loss rate  
            (mass-loss rate of V532 is $\dot{M}=1.6\cdot10^{-5}\rm M_{\odot}/year$).
            This difference is quite normal, because mass-loss rate 
            increases with luminosity and with helium abundance
            \citep{Nugis}.  
            Moreover the nitrogen abundance N/He is $10^{-3}$ ($\sim 2.7$ solar) 
            unlike V532, where N/He is $3.0 \times 10^{-3}$ ($\sim 7.1$ solar).  

\begin{table*}\centering
\caption{Derived properties of FSZ35 and V532 in the minimum of 
        brightness (October 2007), and a comparison with similar stars in Milky Way (MW). 
        $X_H$ is mass fraction of hydrogen} 
\label{tab:parmodel}
\bigskip
\begin{tabular}{lccccccccc}
\hline
Star   & Gal. & Sp.   & $T_*$ &   $R_*$     &$\log L_*$     & $\log \dot{M}_{\rm uncl}$    & $v_{\infty}$  & $X_H$ & Ref       \\
       &      & type  & [kK]  &[$R_{\odot}$]& [ $L_{\odot}$]& [$\rm M_{\odot}/year$]  & [km/s]        & [\%]  &            \\
\hline
WR124   & MW   &  WN8h& 32.7  &    18.0  &  5.53           &     -4.2                &  710          & 13     & [1] \\

WR116  & MW   &  WN8h & 39.8  &    21.0  &   6.0           &     -3.7                & 800           & 10     & [2]  \\

WR89   & MW   &  WN8h & 39.8  &    26.5  &   6.2           &     -4.2                & 1600          & 20     & [2]  \\

WR40   & MW   &  WN8h & 45.0  &    10.6  &   5.61          &     -4.                 &  840          & 15     & [3] \\   
  
WR16   & MW   &  WN8h & 41.7  &    12.3  &   5.68          &      -4.3               &  650          & 23     & [3] \\
       &      &       &       &          &                 &                         &               &        &     \\
 FSZ35 &  M33 &  WN8h &  36.5 &    19    &   5.76          &     -4.1                &  800          &  16.5  &     \\ 
 V532  &  M33 &  WN8h &  34.0 &    20.8  &   5.7           &     -4.3                &  360          &  24    & [4]  \\  

\hline
\multicolumn{10}{l}{\footnotesize [1]- \citet{Crowther99}, [2]- \citet{Hamann},}\\ 
\multicolumn{10}{l}{\footnotesize   [3]- \citet{HHH}, [4]-  \citet{v532model} }
\end{tabular}
\end{table*}


\section{Discussion}\label{sec:disc}

\subsection{Nebular Contribution}\label{sec:disc:neb}

             Although the spectra of V532 and FSZ35 are very similar, 
             a close look at the subtle differences can be illuminating. 
             All the forbidden lines, including [Fe\,{\sevensize III}]$\lambda\lambda
             4658.1,4701.5,5270.4$, are either absent or much
             fainter for FSZ35.
             First we suggested the line near the He\,{\sevensize I}$\lambda$5015 emission belongs
             to the N\,{\sevensize II}$\lambda\lambda 4987-5007$ blend but we did not find reasonable
             solution that could account for the whole nitrogen spectrum
             (N\,{\sevensize II}$\lambda\lambda 5001-10$, N\,{\sevensize III}$\lambda\lambda 4634,40$ 
             and N\,{\sevensize IV}$\lambda 4057$). The line
             position is consistent with the [O\,{\sevensize III}]$\lambda$5007 emission. Total flux
             in this line is $F($[O\,{\sevensize III}]$)\simeq (2.8\pm 0.2)\times10^{-15}\rm erg\,
             cm^{-2}s^{-1}$.  The other component of the doublet,
             [O\,{\sevensize III}]$\lambda$4959, should thus have the flux of
             $\sim 9.3\times10^{-16}\rm erg\, cm^{-2}s^{-1}$ and is unreachable in the
             spectrum. 

             Using the best-fit CMFGEN model spectrum as input, we run a set of
             Cloudy \citep{cloudy} version 08.00
             photoionization models. We restrict the size of
             the putative nebula around FSZ35 with the outer radius of 0.5\,pc
             (there are no indications for resolved diffuse nebular emission around
             the star) and fix the inner radius to 0.1\,pc. Metallicity pattern
             {\tt H\,{\sevensize II}  region} was used with the metallicity decreased by a factor of
             2. We do not change nitrogen and helium abundances in the nebula
             because they can be less affected by chemical evolution of the
             star and represent rather the chemical composition of its early
             hydrogen-rich ejecta. 

\begin{figure}
\centering
\epsfig{file=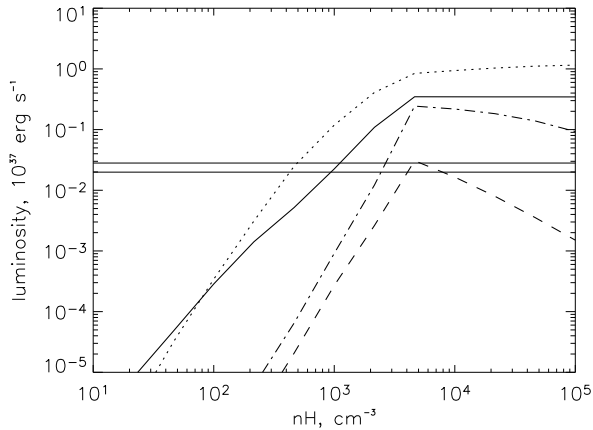,width =1.0\columnwidth}
\caption{Dependence of nebular line luminosities on the hydrogen
  density of the putative nebula around FSZ35. Solid is H$\beta$,
  dotted is [O\,{\sevensize III}]$\lambda$5007, dashed is [S\,{\sevensize II}]$\lambda$6717 and
  [N\,{\sevensize II}]$\lambda$6583 is shown by a dash-dotted line. Horizontal lines
  show the measured luminosity of the [O\,{\sevensize III}] line (1$\sigma$
  confidence interval).}
\label{fig:lines}
\end{figure}

             The only varied parameter was hydrogen density. For several values of
             this parameter, we calculate the luminosities of several nebular lines
             and give them in figure \ref{fig:lines}. For comparison, the
             observational estimate for the luminosity in [O\,{\sevensize III}]$\lambda$5007 is
             given. Evidently, the observed value is in good agreement with the
             predicted nebular luminosity for hydrogen density $n_H \sim 300\rm cm^{-3}$. Note
             however that oxygen abundance may be lower in the ejecta consisting the
             nebula (see below). If oxygen is under-abundant by a factor of
             several, the observed flux is consistent with $n_H \sim 10^3 \rm cm^{-3}$.

             We also expect emission lines of [S\,{\sevensize II}] and [N\,{\sevensize II}] in the red
             spectral range to be much fainter. However, they may
             be stronger if the matter in the nebula is denser and chemically
             evolved, or shock excitation also contributes to the
             emission of the nebula.

             Relying on the similarity of the central stars, one
             may propose that 
             the nebula is similar to M1-67 observed around WR124 \citep{Crowther99}. 
             Strongest nebular lines in the spectrum of M1-67 are lower-excitation lines
             such as [S\,{\sevensize II}]$\lambda\lambda 6717,31$\AA, 
             [N\,{\sevensize II}]$\lambda\lambda 6548,84$\AA \ and
             [O\,{\sevensize II}]$\lambda3727$\AA. 

             Another effect making the putative Wolf-Rayet nebula of FSZ35 hard to
             detect with the data on hand may be oxygen
             under-abundance (by a factor of 5$\div$10) reported for M1-67 and other
             Wolf-Rayet nebulae and related objects (see \citet{Esteban} and
             references therein). Oxygen deficit is a predicted
             by-product of the CNO
             cycle and should have (as well as carbon and neon deficits) an
             amplitude similar to that of nitrogen over-abundance. 

\subsection{Lack of variability}\label{sec:disc:var}

             The spectrum analysed in this work is identical within
             one spectral sub-class to the spectrum published by \citet{massey98}. 
             We also classify FSZ35 as WN8. Unlike V532, FSZ35 does
             not demonstrate any prominent spectral variability. 
            
             It is also fairly stable photometrically. 
             {\it V}-band magnitude of the object was 18$\mag{\ .}$7 in 1993 \citep{massey98} and 18$\mag{\ .}$8 in
             2004 \citep{Hartmann06}, that significantly restricts the
             possible amplitude of photometrical variability. In the {\it V} band,
             $\Delta m \lesssim 0\mag{\ .}1$. The lack of photometrical variability
             may be ascribed to the more advanced evolutionary status of
             FSZ35 \citep{meynet2011}. 

\subsection{FSZ35 as a massive runaway}\label{sec:disc:ra}

             FSZ35 is located  at the distance of about
             $35\arcsec{\,}$\ ($\sim115$ pc) from the
             association OB~128. 
             Suppose that once FSZ35 was a member of the association and 
             was ejected via slingshot-type dynamical interaction. 
             If its peculiar velocity is  $\sim 100\kms$, as for WR124 \citep{wr124vel,marchenko}, 
             it could have been expelled from the parent cluster about
             a million years ago.

             Offset positions with respect to the probable parent associations
             (at distances $\sim 100\pc$) and unexpectedly large peculiar
             velocities (of the order $\sim 100\kms$) seem to be common for
             very luminous and massive stars like V532, FSZ35 and Galactic late WN
             stars like WR20a and WR124. A scenario was proposed by \citet{GG11}  
             that applies three-body dynamical interaction in the
             cores of young massive
             clusters and star-forming regions (similar to 30~Doradus) to reproduce the
             observed population of massive runaways. This scenario has the
             disadvantage of relying heavily on the formation of massive stars in
             the cores of massive young clusters that are, in contrast
             with LMC, absent in M33, where massive stars are rather formed
             in dispersed stellar associations.

             Instead we would rather propose that very massive stars are
             formed in dense groups containing several stars each. This is
             confirmed, for example, by the multiplicity increasing
             with stellar mass \citep{Zinnecker} both in young star
             clusters and associations. It is reasonable
             that higher fraction of massive binaries will be
             accompanied by a higher fraction of massive multiple
             systems. When formed, such systems are often unstable \citep{kiseleva}
             and dynamic interaction between its
             components should both produce a larger fraction of runaways at these
             masses ($\sim 100\Msun$) and a larger fraction of
             binaries. The characteristic peculiar velocities ($\sim 100\kms$) of these
             ``childhood runaways'' may be reproduced if the initial
             spatial sizes of the systems are $\lesssim 10^{14}\rm cm$.

\section{Conclusions}\label{sec:con}

        We analyse the optical spectrum of the little-studied WNL star FSZ35 in M33. 
        About 40 spectral lines in the $4000\div 5500$\AA\ wavelength range are identified. 
        We classify FSZ35 as a WN8 star, confirming the result of \citet{massey98} 
        and put upper limits for the spectral and photometrical
        variability of this object.

        Using non-LTE code CMFGEN we estimate the parameters of FSZ35
        (bolometric luminosity, stellar radius, mass loss rate, wind velocity, elementary
        abundances) and compare them to the corresponding parameters of other WN8 stars 
        including the LBV star V532 during the minimum of brightness. 
        FSZ35 is a H-rich WN8 star where the mass fraction of hydrogen is 17\% (H/He=0.8). 
        The best-fit parameters of the model are: luminosity $L=5.3\cdot 10^5
        L_\odot$, mass loss rate $2.1\cdot 10^{-5} {M}_\odot/\mbox{year}$,
        nitrogen abundance N/He =13.5(N/He)$_{\odot}$, effective temperature at
        hydrostatic radius $T_{*}=36\,470 \rm K$ ($R_*=18R_\odot$) and at
        the Rosseland photosphere
        $T_{tau=2/3}=35\,160 \rm K$. Derived parameters of FSZ35 atmosphere correspond to a
        typical WN8 star.

        We find that FSZ35 has a surrounding nebula, possibly  of low excitation
        and deficient in oxygen, that decreases its detectability.

        Position of FSZ35 at the outskirts of association 128~OB 
        suggests that it was expelled from this association about a million 
        years ago at a velocity of $\sim 100 \kms$.
 
\acknowledgements
        This work makes use of the data taken from the public archive of the
        Special Astrophysical Observatory. 
        We wish to thank John D. Hillier for his great code CMFGEN, both
        comprehensive and user-friendly, that we applied to fit and analyse
        the data.  
        We also would like to thank the referee
        Wolf-Rainer Hamann for valuable comments. 
        One of us (P. A.) thanks leading
        scientific schools grant NSh-7179.2010.2 for support.


\end{document}